\documentclass[12pt]{iopart}
\usepackage{iopams}
\usepackage{graphicx}
\usepackage[usenames,dvipsnames]{color}
\input{epsf}
\newcommand{\trm}{\mathfrak{t}}
\newcommand{\rfl}{\mathfrak{r}}
\newcommand{\avtrm}{\widehat{\trm}}
\newcommand{\avrfl}{\widehat{\rfl}}
\begin{document}

\title[Non-homogeneous persistent random walks and L\'evy-Lorentz gas]{Non-homogeneous persistent random walks and averaged environment for the L\'evy-Lorentz gas}

\author{Roberto Artuso$^{1,2}$, Giampaolo Cristadoro$^3$, Manuele Onofri$^1$ and Mattia Radice$^{1,2}$}

\address{$^1$ Dipartimento di Scienza e Alta Tecnologia and Center for Nonlinear and Complex Systems, Universit\`a degli Studi dell'Insubria, Via Valleggio 11, 22100 Como Italy}
\address{$^2$ I.N.F.N. Sezione di Milano, Via Celoria 16, 20133 Milano, Italy}
\address{$^3$ Dipartimento di Matematica e Applicazioni, Universit\`a degli Studi di Milano-Bicocca, Via Cozzi 55, 20125 Milano, Italy}
\eads{\mailto{roberto.artuso@uninsubria.it}, \mailto{giampaolo.cristadoro@unimib.it}, \\\mailto{m.onofri@hotmail.com}, \mailto{m.radice1@uninsubria.it }}
\vspace{10pt}

\begin{abstract}
We consider transport properties for a non-homogeneous persistent random walk, that may be viewed as a mean-field version of the L\'evy-Lorentz gas, namely a 1-d model characterized by a  fat polynomial tail of the distribution of scatterers' distance, with parameter $\alpha$. By varying the value of $\alpha$  we have a transition from normal transport to superdiffusion, which we characterize by appropriate continuum limits. 
\end{abstract}

\vspace{2pc}
%\submitto{\JSTAT}
%\maketitle

\section{Introduction}
Persistent random walks were first introduced by F\"urth and Taylor \cite{Fu-p,Ta-p} almost a century ago, as a model for particle diffusion by discontinuous movements: in their simplest version they consist of random walkers with nearest neighbour jumps, where the walker has a probability $\trm$ of jumping in the same direction and a probability $\rfl=1-\trm$ of reversing the direction of motion. Even in this quite simple formulation we perceive that what makes such walks nontrivial is the presence of correlated steps. The basic properties of the time evolution of a localized initial distribution were derived in different ways: see for instance \cite{Go-p,Ka-p,RH-p,We-p}, to which we refer the reader for a more detailed early bibliography. A few results will be mentioned in the following sections. \\
Persistent random walks have been recognized as a natural model for a number of relevant settings, from long-chain polymers \cite{Pa-a}, to chemotaxis \cite{Sc-a}, to active matter \cite{De-a,den}. Many of the associated statistical properties remain largely unexplored, particularly when homogeneity is violated: this comes to no surprise, since, even in standard random walks few results are known when transition probabilities do not share translational invariance of the lattice (see the discussion in \cite{B:Hu}).\\
A particular  persistent random walk model that attracted recently much interest is the so called L\'evy-Lorentz gas, originally formulated in \cite{BFK-ll}: the lattice of integers is  populated  by randomly placed scatterers separated by distances whose probability distribution decays, for large separations, with a fat (L\'evy) tail
\begin{equation}
\label{dprob}
\mu(\xi)\sim \xi^{-(1+\alpha)}.
\end{equation}
In the {\it quenched} (non-equilibrium -see next section for details-) version \cite{BCLL-ll,ST-ll} a single (typical) realization of the distribution of  scatterers is considered. Here one is interested in the distributions of particles (initially concentrated at the origin), assuming that they propagate with constant velocity, unless they reach a scatterer, where they are reflected or transmitted with equal probabilities. This amounts to assign a transmission coefficient $\trm=1/2$ to all sites where a scatterer is present, and $\trm=1$ at otherwise. The corresponding {\it annealed} model \cite{BFK-ll,BCV-ll} considers then the probability distribution of particles averaged over disorder realizations (positions of the scatterers). 
\\
We introduce in this paper a 1d model of a non-homogeneous persistent random walk that arises when we consider an {\it averaged} L\'evy-Lorentz gas: namely we assign a distribution of local reflection and transmission coefficients obtained (prior to let particles evolve) by an average over distribution of scatterers (such an average yields nontrivial, non translational invariant, coefficients for configurations where a scatterer is always placed at the origin). The result will be a non-homogeneous persistent random walk, whose transport properties are the issues at stake in our work. 
In particular we will show, with the help of a continuum approximation,  that a non-trivial anomalous diffusive regime  appear below $\alpha =1$:
\begin{equation}
\langle x^2_t \rangle \sim \left\{
\begin{array}{ll}  t^{\frac{2}{1+\alpha}} \qquad & 0 < \alpha <1\\
t \qquad & 1< \alpha < 2
\end{array} \right.
\end{equation}
We remark that besides theoretical interest, and the wealth of applications that persistent walks exhibit, a further motivation to investigate such models comes from the experimental realizations (in 3d) of a very closely related system, the so called L\'evy glass \cite{BBW-ll}. \\
The paper is organized as follows: in section \ref{s:LL} we briefly review the L\'evy-Lorentz gas, with a focus on transport properties, and then we introduce the persistent random walk that is the main object of our paper; in section \ref{s:TP} we then discuss normal and anomalous transport properties that our model exhibits for different values of $\alpha$; finally in section \ref{s:CP} we summarize our findings.

\section{\label{s:LL}1d L\'evy-Lorentz gas and the averaged model}

\subsection{The L\'evy-Lorentz gas}

As we already mentioned the model is defined in two steps: first we distribute scatterers on a one dimensional lattice, separating them with distances chosen independently according to a fat tailed probability distribution
\begin{equation}
\label{dist-p}
\mu(n)=\frac 1{\zeta(1+\alpha)\cdot n^{(1+\alpha)}} \quad \quad  n \ge 1
\end{equation}
where $\alpha \in (0,2)$; notice that in this range the variance of the distance between neighbouring scattering sites diverges, while in the restricted range $0<\alpha<1$ also the first moment is infinite.
Then the stochastic dynamics is determined by choosing a starting site, and let walkers (with unit absolute velocity) initially jump to the right or the left with equal probability. 
At later times each walker maintains the direction of motion until it finds a scatterer: at such events direction is preserved or reversed with equal probability 1/2. 
\\
In the {\it quenched case}  very few results have been established, notably the central limit theorem in the range $1<\alpha <2$ \cite{BCLL-ll}: normal diffusion is not rigorously proven, yet no indication of possible anomalies emerges. A different scenario has been proposed in the {\it annealed} case where it has been suggested \cite{BCV-ll} that the second moment grows linearly in time only for $\alpha>3/2$, while for smaller values of the exponent the behaviour is expected to be
\begin{equation}
\label{2mom:BCV}
\langle x^2_t \rangle \sim \left\{
\begin{array}{ll} t^{\frac52-\alpha} \qquad & 1\leq \alpha \leq 3/2 \\
t^{\frac{2+2\alpha -\alpha^2}{1+\alpha}} \qquad & 0 < \alpha <1.
\end{array} \right.
\end{equation}
The key ingredient in deriving such expression is a decomposition of the propagator into a scaling part and a contribution for very long jumps
\begin{equation}
\label{prob:BCV}
p(r,t)=\frac 1 {\ell(t)}{\cal F}\left( \frac r {\ell(t)} \right) + {\cal H}(r,t).
\end{equation}
The scale $\ell(t)$ and the long jump term ${\cal H}$ (whose space integral vanish in the long time limit) are determined by using estimates for the associated resistance model \cite{BGA-ll}: such a decomposition of the probability distribution has been recently considered in a wider context \cite{bigj}.

\subsection{The L\'evy-Lorentz gas as a persistent random walk}

A persistent random walk on a one dimensional lattice is defined in terms of the quantities $\trm_j$ and $\rfl_j$, that determine for each site the probability of being transmitted or reflected: in order to write down the evolution of the probability distribution of the walker, it is convenient to split it according to the direction of motion in the following way
\[
\begin{array}{l}
R_j(n)= \mbox{Prob}\,(\mbox{\small{~The walker is at site {\it j} after {\it n} steps and leaves to the right~}})\\
L_j(n)=\mbox{Prob}\,(\mbox{\small{~The walker is at site {\it j} after {\it n} steps and leaves to the left~}}).\end{array}
\] 
Here we are adopting the notation of \cite{RH-p} where the forward Kolmogorov equations are written as
\begin{equation}
\label{fKolm}
\begin{array}{lll}
R_j(n+1)&=&\trm_j\cdot R_{j-1}(n)+ \rfl_j \cdot L_{j+1}(n)\\
L_j(n+1)&=&\trm_j\cdot L_{j+1}(n)+ \rfl_j \cdot R_{j-1}(n),
\end{array}
\end{equation}
with the  choice of  initial conditions:
\begin{equation}
\label{fKin}
R_0(0)=L_0(0)=\frac 12, \quad R_j(0)=L_j(0)=0 \,\,\,\,\forall j \neq 0.
\end{equation}
Given a  realization $\Omega$ of the  environment, the quenched L\'evy-Lorentz gas consists in assigning 
\begin{equation}
\label{qtr}
\rfl _j=\emph{r}\cdot\delta_{\Omega}(j)\qquad \trm_j=1-\rfl_j
,
\end{equation}
where $\delta_{\Omega}(j)=1$ if in the realization $\Omega$ the site $j$ is occupied by a scatterer, and $\delta_{\Omega}(j)=0$ otherwise. In the following, we choose $r=1/2$ for the sake of simplicity. Note that, as we deal with  {\it nonequilibrium} case only, for any realization  $\delta_{\Omega}(0)=1$. \\
The probability for a walker to be at site $j$ after $n$ steps will consequently be
\begin{equation}
\label{perP}
P_j(n)=R_j(n)+L_j(n),
\end{equation}
while
\begin{equation}
\label{perJ}
M_j(n)=R_j(n)-L_j(n)
\end{equation}
gives the (rightwise) current at time $n$.
The annealed version of the model consists in evolving eq. (\ref{fKolm}) for a single realization, and then averaging probabilities over the different environments; the model we introduce in this paper reverses the two operations (which do not commute): we first average the different environments, and then we study the evolution over such an averaged landscape:
in a sense this amounts to consider mean field evolution over a fast changing environment.

\subsection{The averaged model}

By averaging Eq.~(\ref{qtr}) over realizations,
we get:
\begin{equation}
\label{avqres}
\avrfl_j= \frac 1 2 \varpi_j,
\end{equation}
where $\varpi_j$ is the probability of finding a scatterer at the position $j$ (under the condition that a scatterer is placed at the origin).
This means (in terms of the distance probability (\ref{dist-p})):
\begin{eqnarray}
\label{varp-n}
\varpi_j&=&\mu(j)+\sum_{k_1+k_2=j}\mu(k_1)\mu(k_2)+ \cdots + \mu(1)^j  \quad \textrm{for} \, j\ge 1\\
\varpi_0&=&1
\end{eqnarray}
In order to get the asymptotic behaviour, it is as usual convenient to introduce the generating function:
\begin{equation}
\label{varp-gen}
%\mathbf{G}(z)=\sum_{m=0}^\infty\,\varpi_m z^m=\sum_{m=0}^\infty z^m\sum_{k=1}^m\sum_{l_1+l_2 + \cdots +l_k=m}\mu(l_1)\mu(\l_2)\cdots\mu(l_k),
\mathbf{G}(z)=\sum_{m=0}^\infty\,\varpi_m z^m=1+\sum_{m=1}^\infty z^m\sum_{k=1}^m\sum_{l_1+l_2 + \cdots +l_k=m}\mu(l_1)\mu(\l_2)\cdots\mu(l_k),
\end{equation}
this easily leads to the expression
\begin{equation}
\label{varp-gen2}
\mathbf{G}(z)=\frac{1}{1-{\cal G}_\alpha(z)},
\end{equation}
where, by (\ref{dist-p})
\begin{equation}
\label{calg}
{\cal G}_\alpha(z)=\sum_{n=1}^\infty \mu(n)z^n=\frac{1}{\zeta(1+\alpha)}\mathrm{Li}_{1+\alpha}(z), 
\end{equation}
where $\mathrm{Li}_s$ denotes the polylogarithms \cite{poll}.
The asymptotic form of $\varpi_m$ is thus estimated by Tauberian theorems for power series \cite{Feller}, once we take into account the expression of polylogarithms close to $z=1^-$ \cite{Fla}: the leading order is
	\begin{equation}\label{loflag}
	\mathcal{G}_{\alpha}(z)\sim
	\left\{
	\begin{array}{ll}
	1+\frac{\Gamma(-\alpha)}{\zeta(1+\alpha)}(1-z)^{\alpha} \quad& \mathrm{for}\ 0<\alpha<1 \\
%	1-\frac{6}{\pi^2}(1-z)\log\left(\frac{1}{1-z}\right)   \quad& \mathrm{for}\ \alpha=1\\
	1-\frac{\zeta(\alpha)}{\zeta(1+\alpha)}(1-z) \quad& \mathrm{for}\ 1<\alpha < 2.
	\end{array}\right.
	\end{equation}
%while for $\alpha=1$ we have:
%	\begin{equation}
%	\mathcal{G}_{\alpha}(z)\sim 1-\frac{6}{\pi^2}(1-z)\log\left(\frac{1}{1-z}\right).
%	\end{equation}
The corresponding asymptotic values of having a scatterer at site $n$ are	
	\begin{equation}\label{aspro}
		\varpi_n\sim \pi_n=\left\{
		\begin{array}{ll}
			\frac{\alpha\sin\left(\pi\alpha\right)}{\pi}\frac{\zeta\left(1+\alpha\right)}{n^{1-\alpha}} \quad& 0<\alpha<1 \\
%			\frac{\pi^2}{6}\frac{1}{\log n} \quad& \alpha=1 \\
			\frac{\zeta\left(1+\alpha\right)}{\zeta\left(\alpha\right)} \quad& 1<\alpha < 2.
		\end{array}\right.
	\end{equation} 
For the rest of the paper the {\it averaged} model will be the persistent random walk with reflection coefficients $\avrfl_j=1/2\cdot \pi_j$, with $\pi_j$ given by (\ref{aspro}) for any value of $j$. 
The corresponding  Kolmogorov equations are thus
\begin{equation}
\label{fKolmav}
\begin{array}{lll}
R_j(n+1)&=&\avtrm_j\cdot R_{j-1}(n)+ \avrfl_j \cdot L_{j+1}(n)\\
L_j(n+1)&=&\avtrm_j\cdot L_{j+1}(n)+ \avrfl_j \cdot R_{j-1}(n).
\end{array}
\end{equation}
We notice, for further reference, that in the finite average $1<\alpha < 2$ regime the coefficients are indeed constants. 
Conversely, the situation is highly non trivial for $0<\alpha < 1$, since in this case we have an effective space dependence of the reflection probabilities: even ordinary random walks with jumping rates that break translational invariance are known to be quite difficult to study \cite{B:Hu,Ginn}.
	
\section{\label{s:TP}Transport properties of the averaged model}

\subsection{The continuum limit}
The continuum limit has been considered in many of the classical papers, as \cite{Go-p,Ka-p,RH-p,We-p}, our approach is close to \cite{RH-p}: we
let $x=n\cdot \delta x$ and $t=m \cdot \delta t$, and Taylor expand (\ref{fKolmav}) up to second order:
\begin{equation}
\label{fKcont}
\left\{
\begin{array}{ll}
R +\dot{R}\delta t+\frac 12 \ddot{R} \delta t^2 &= \avtrm R-\avtrm R'\delta x+ \frac {1}{2}\avtrm R''\delta x ^2 +\avrfl L+\avrfl L' \delta x+\frac{1}{2}\avrfl L'' \delta x^2\\
L +\dot{L}\delta t+\frac 12 \ddot{L} \delta t^2 &= \avrfl R-\avrfl R' \delta x + \frac 1 2 \avrfl R''\delta x^2 +\avtrm L +\avtrm L' \delta x +\frac 12 \avtrm L'' \delta x^2,
\end{array}
\right.
\end{equation}
where $R=R(x,t),\,\,L=L(x,t)$, $\avtrm=\avtrm(x),\,\,\avrfl=\avrfl(x)$, the dot stands for time derivative, while the prime indicates spatial derivatives.
Now we consider the total probability and the flux
\begin{equation}
\label{proflu}
\begin{array}{rll}
P(x,t)&=&R(x,t)+L(x,t) \\
M(x,t)&=&R(x,t)-L(x,t),
\end{array}
\end{equation}
and, by adding and subtracting the identities (\ref{fKcont}) we get
\begin{equation}
\label{PMcont}
\left\{
\begin{array}{rl}
\dot P \delta t + \frac 1 2 \ddot{P} \delta t^2&= -M' \delta x +\frac 1 2 P'' \delta x^2\\
M+\dot{M} \delta t +\frac 1 2 \ddot{P} \delta t^2&= (\avtrm -\avrfl)M-(\avtrm-\avrfl)P' \delta x +\frac 1 2 (\avtrm -\avrfl)M'' \delta x ^2.
\end{array}
\right.
\end{equation}

In order to get a closed equation for $P$, we have to specify a scaling limit for the former identities. Since our focus is on the asymptotic regime, as we want to estimate the second moment in the long time limit, we employ the diffusion approximation \cite{RH-p,We-p}, where $\delta x,\, \delta t \to 0$ by keeping $D_0=\delta x^2/\delta t$ constant (we will consider $D_0=1$ in what follows).
From the second of (\ref{PMcont}), we obtain
\begin{equation}
\label{subM}
M' \delta x=-\frac{\partial}{\partial x}\left[\left(
\frac{1-2\avrfl}{2\avrfl}\right)\frac{\partial P}{\partial x}
\right]\delta x^2+ O(\delta x^3),
\end{equation}
and by substituting it into the first of (\ref{PMcont}) we obtain  a  closed equation for $P$ %is written as
\begin{equation}
\label{Pandiff}
\frac{\partial P}{\partial t} =\frac{1}{2}\frac{\partial}{\partial x} \left( \Theta_{\alpha}(x) \frac {\partial P}{\partial x} \right)
\end{equation}
with
\begin{equation}
\label{diffx}
\Theta_{\alpha}(x)=\frac{\avtrm (x)}{\avrfl (x)}=
\left\{
\begin{array}{ll}
2\frac{\zeta (\alpha)}{\zeta(1+\alpha)}-1\qquad & 1 < \alpha < 2 \\
\frac{1-\gamma/ |x|^{1-\alpha}}{\gamma /|x|^{1-\alpha}}\qquad &0 < \alpha <1,
\end{array}
\right.
\end{equation}
where $\gamma=\alpha\sin(\pi \alpha)\zeta(1+\alpha)/2\pi$ (see (\ref{aspro})).\\
We remark that there is another scaling limit in which the tail of the distribution accounts for the ballistic peaks we will comment upon in the last section: as a matter of fact if we put $\avtrm =1-\delta t/(2\tau)$ and we let $\delta x,\,\,\delta t \to 0$ by keeping constant $\delta x/\delta t=c$
we get (here we report the result only in the simplified case in which $\tau$ is constant) the telegrapher's equation
\begin{equation}
\label{con-tel}
\frac{\partial^2 P}{\partial t^2}+\frac{1}{\tau}\frac{\partial P}{\partial t}=c^2\frac{\partial^2 P}{\partial x^2},
\end{equation}
as observed and discussed by many authors \cite{Go-p,Ka-p,RH-p,We-p} (see also \cite{IPZ} for a discussion of the relative role of different time scales in the telegrapher's equation).
\subsubsection{Second moment asymptotics\\}
Given the diffusion equation Eq.(\ref{Pandiff}) it is possible to derive the asymptotic behaviour of the second moment.\\ 
\begin{figure}[h!]
\centering
\includegraphics[width=13cm]{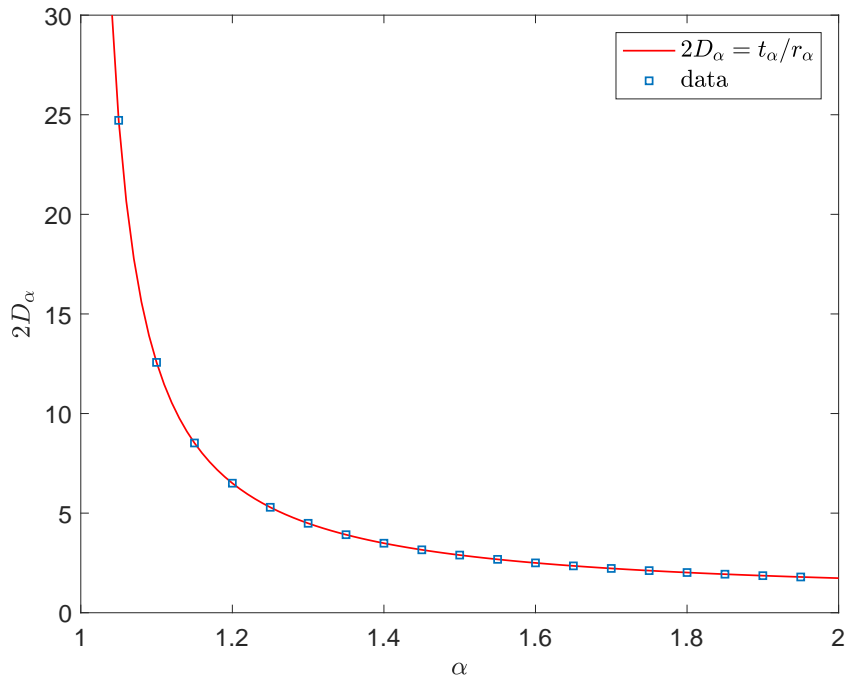}
\vskip4mm
\caption{\label{diff-n-fig}Slope of linear growth of the second moment as obtained by numerically evolving the forward Kolmogorov equations (\ref{fKolmav}) (squares)  and the analytic prediction in terms of the diffusion constant (\ref{diffx}). Each numerical slope has been obtained by evolving the system up to time $2^{15}$.}
\end{figure} 
We observe that, for  for $1<\alpha<2$  (where a central limit theorem holds for the quenched L\'evy-Lorentz case \cite{BCLL-ll}, and the average distance between scatterers is finite), we have a persistent random walk with constant transmission and reflection coefficients and consistently 
 the diffusion coefficient $\Theta_{\alpha}/2$ does not depend on space. In this regime we have thus  normal diffusion, with 
\begin{equation}
\label{2norm}
\langle x_t^2 \rangle \sim 2 D_{\alpha} \cdot t,
\end{equation}
where $D_{\alpha}=\avtrm/2\avrfl$, as early deduced in \cite{RH-p}.
The agreement with numerical simulations is indeed excellent, see fig. ({\ref{diff-n-fig}).
\\
The interesting regime is obviously the case when $\alpha \in (0,1)$, as  $\avtrm(x)$ and $\avrfl(x)$ are not constant. This is reflected into a space-dependent diffusion coefficient, that  decays algebraically with the distance to the origin. 
\begin{figure}[h!]
\centering
\includegraphics[width=13cm]{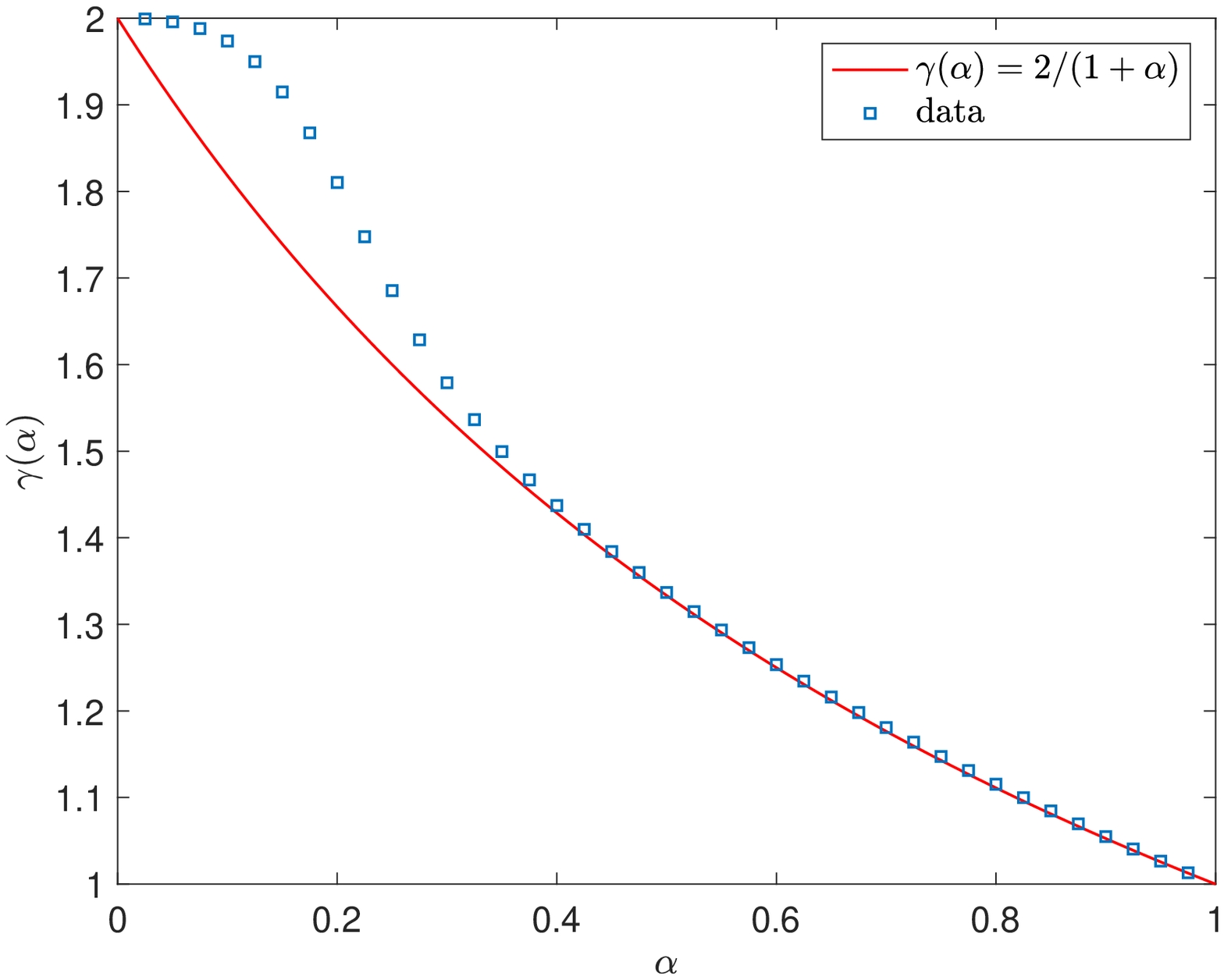}
\vskip4mm
\caption{\label{diff-a-fig}Asymptotic growth exponent of the second moment $\langle x^2_t\rangle \sim t^{\gamma(\alpha)}$: finite time estimate deviate from theoretical prediction when $\alpha$ is close to $0$: see section (\ref{sub:err}) for a discussion about this point. Each numerical exponent has been obtained by evolving the system up to time $2^{18}$.}
\end{figure}
In this regime,  the  solution of the diffusion equation (\ref{Pandiff})
may be written as \cite{HPdiff,OPdiff,RGF} (see also \cite{CC1,CC2} for further discussion):
\begin{equation}
\label{solDx}
P(x,t)=\frac{\left[(1+\alpha)^{1-\alpha}\Lambda t\right]^{-1/(1+\alpha)}}{2\Gamma(1/(1+\alpha))}\exp \left[\frac{-|x|^{1+\alpha}}{(1+\alpha)^2 \Lambda t}\right],
\end{equation}
where we have considered the leading order of (\ref{diffx}) at large distances, and $\Lambda=1/(2\gamma)$.
From (\ref{solDx}) we see that the diffusion is anomalous as  the second moment is:
\begin{equation}
\label{an2mom}
\langle x^2_t \rangle =\, C_{\alpha}\cdot  t^{2/(1+\alpha)},
\end{equation}
where $C_{\alpha}=\frac{\Gamma(3/(1+\alpha))}{\Gamma (1/(1+\alpha))} \left[ (1+\alpha)^2\Lambda \right]^{2/(1+\alpha)}$.
\\
\\ Finally we note explicitly  that the probability distribution is of the form (see (\ref{prob:BCV}))
\begin{equation}
\label{scalP}
P(x,t)=\frac{1}{\ell(t)} {\cal F}\left(\frac{|x|}{\ell(t)}\right)
\end{equation}
where
\begin{equation}
\label{callL}
\ell(t)= t^{1/(1+\alpha)},
\end{equation}
and ${\cal F}(y)$ is a (stretched) exponential.
The result is consistent with the scaling predicted in \cite{BCV-ll}, while the form of the scaling function ${\cal F}$ is different, as in the present case it cannot contribute anomalies, due to its fast decay to zero for large arguments. Notice that in different situations, with slower decay of the scaling function, the moments may not be determined by the scaling (\ref{callL}), see \cite{SZS,CDP}. In the present case, as discussed in the next section, also ballistic peaks cannot influence the asymptotic growth of high order moments, so our model is characterized by weak anomalous diffusion \cite{Vulp}, which means that there is a single scale ruling the behaviour of the whole moments spectrum
\begin{equation}
\label{momsp}
\langle |x_t|^q\rangle \sim t^{q/(1+\alpha)}.
\end{equation}

\subsection{\label{sub:err}Ballistic peaks and finite-time estimates}
One feature that is not captured by the diffusion approximation, is the structure of the tails of the propagating probability distribution: as a matter of fact in the discrete setting, with initial conditions given by (\ref{fKin}), $P_j(n)$ is zero for any position $j>n$, and the contribution to the front is given by walkers whose velocity never reversed up to time $n$. In the present case - differently from the annealed L\'evy-Lorentz model \cite{BFK-ll,BCV-ll} -  such peaks do not contribute to the asymptotic behaviour of the second moment, while they may influence intermediate time estimates, especially as $\alpha$ approaches $0$, as witnessed by fig. (\ref{diff-a-fig}).\\
\begin{figure}[h!]
\centering
\includegraphics[width=13cm]{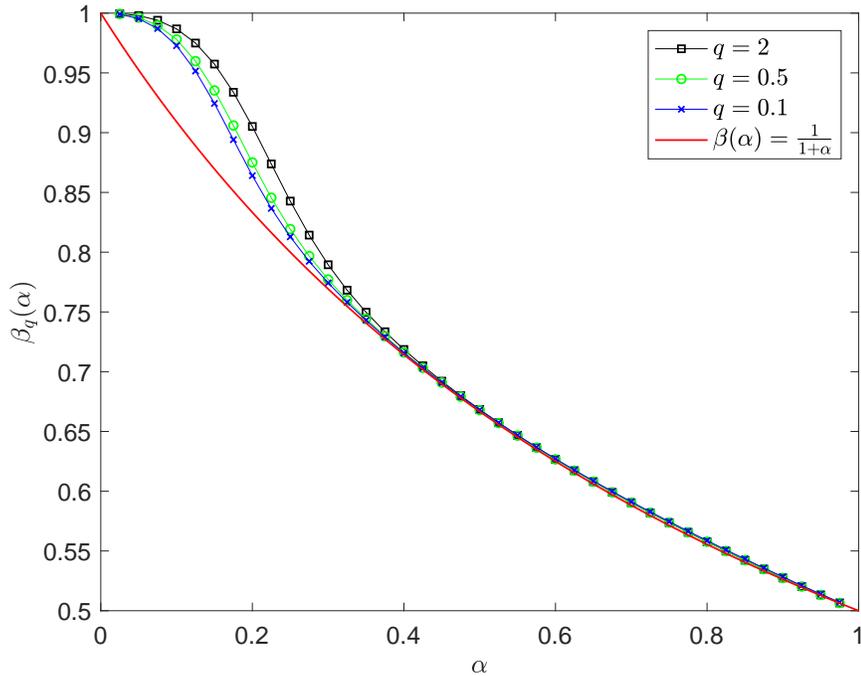}
\vskip4mm
\caption{\label{diff-q-fig}Asymptotic growth exponent of q-th order moment $\langle |x_t|^q\rangle \sim t^{q\cdot \beta_q(\alpha)}$ for some (low) values of q: low order moments are less influenced by finite time effects.}
\end{figure}
The ballistic peak amplitude can be computed directly from the discrete model: in range $1<\alpha <2$
\begin{equation}
\label{balp-n}
P_n(n)=\frac{1}{2}\avtrm^n,
\end{equation}
which immediately yields a purely exponential decay.

In the anomalous regime $0 < \alpha <1$,
the probability for a walker to be at $j=n$ at time $n$ is 
\begin{equation}
\label{balln}
P_n(n)=\frac{1}{2}\prod_{k=1}^{n-1}\left(1-\frac{1}{2\Lambda k^{1-\alpha}}\right),
\end{equation}
which, for large values of $n$ can be estimated as
\begin{equation}
\label{Nballn}
P_b(n)=P_n(n) \sim \frac{1}{2} \exp \left( -\frac{1}{2 \Lambda \alpha}n^{\alpha} \right)\cdot\exp \left(\frac{1}{2\Lambda \alpha}\right).
\end{equation}
Such a contribution cannot modify the moments' spectrum, due again to the presence of a stretched exponential decay, but may
be relevant for finite times, for $\alpha$ sufficiently close to zero.
In fig. (\ref{diff-q-fig}) we show how such finite estimates have a smaller influence when low order moments are computed \cite{SZS}, yet computations still remain problematic when $\alpha$ is sufficiently close to $0$. 
This may be qualitatively understood if we consider a combined probability,
joining the diffusive and the ballistic terms:
\begin{equation}
\label{pcomb}
P_{eff}(x,t)=P_b(x,t)+{\cal C}(t)P(x,t),
\end{equation}
where $P$ is given by (\ref{solDx}), while (\ref{Nballn})
\begin{equation}
\label{Xballx}
P_b(x,t)=\frac 1 2 \exp \left( -\frac{t^\alpha}{2\Lambda\alpha}\right)\cdot\left(\delta(x-t)+\delta(x+t)\right),
\end{equation}
while ${\cal C}$ is chosen to have a normalized probability distribution:
\begin{equation}
\label{Cnorm}
{\cal C}(t)=1-\exp\left(-\frac{t^\alpha}{2\Lambda \alpha}\right).
\end{equation}
In order to get a crude estimate of the relevance of ballistic peaks for finite times, we may evaluate the ratio:
\begin{equation}
\label{rat}
\frac{\langle x^2_t \rangle_{\mathrm{bal}}}{\langle x^2_t \rangle_{\mathrm{dif}}}=\frac{\exp\left( -\frac{t^\alpha}{2\Lambda\alpha}\right)t^2}{\left(1-\exp\left(-\frac{t^\alpha}{2\Lambda \alpha}\right)\right)\cdot\frac{\Gamma(3/(1+\alpha))}{\Gamma (1/(1+\alpha))}\left[ (1+\alpha)^2\Lambda t \right]^{2/(1+\alpha)}}.
\end{equation}
At the stopping time of our numerical simulations this ratio is small ($\leq 0.005$), for $0.4 \lesssim \alpha$: to get the same value of (\ref{rat}) for $\alpha=0.1$ we would need around $10^{16}$ iterations.
\section{\label{s:CP}Conclusions}
In this paper we have introduced and studied a non-homogeneous persistent random walk, where reversal probability decreases like a power law with respect to the distance from the starting point. This model may be viewed as a non trivial extension of conventional persistent random walks or as the limiting case of non equilibrium L\'evy-Lorentz gas in a fast changing environment. Two different regimes are singled out, the first characterized by normal transport, being indeed equivalent to a persistent random walk with constant reversal probability, while the second exhibits superdiffusion, with an exponent analytically computed via a suitable continuum limit.

%\end{multicols}

\section*{References}
\begin{thebibliography}{99}

\bibitem{Fu-p} F\"urth R 1920 Die Brownsche Bewegung bei Bercksichtigung einer Persistenz der Bewegungsrichtung. Mit. Anwendungen auf die Bewegung lebender Infusorien {\it Z. f\"ur Physik} {\bf 2} 244
\bibitem{Ta-p} Taylor G I 1922 Diffusion by continuous movements {\it Proc. London Math. Soc.} {\bf 20} 196
\bibitem{Go-p} Goldstein S 1951 On diffusion by discontinuous movements, and on the telegraph equation {\it Quart. J. Mech. Applied. Math.} {\bf IV} 129
\bibitem{Ka-p} Kac M 1974 A stochastic model related to the telegrapher's equation {\it Rocky Mount. J. Math.} {\bf 4} 497
\bibitem{RH-p} Renshaw E and Henderson R 1981 The correlated random walk {\it J. Appl. Prob.} {\bf 18} 403
\bibitem{We-p} Weiss G H 2002 Some applications of persistent random walks and the telegrapher's equation {\it Physica {\rm A}} 311, 381
\bibitem{Pa-a} Patlak C S 1953 Random walk with persistence and external bias {\it Bull. Math. Biophys.} {\bf 15} 311
\bibitem{Sc-a} Schnitzer M J 1993 Theory of continuum random walks and application to chemotaxis {\it Phys. Rev. {\rm E}} {\bf 48} 2553
\bibitem{De-a} Detcheverry F 2017 Generalized run-and-turn motions: from bacteria to L\'evy walks {\it Phys. Rev. {\rm E}} {\bf 96} 012415
\bibitem{den} Escaff D, Toral R, Van den Broeck C and Lindenberg K A continuous-time persistent random walk model for flocking arXiv:1803.02114
\bibitem{B:Hu} Hughes B D 1995 {\it Random Walks and Random Environments. Volume 1: Random Walks} (Oxford: Clarendon Press)
\bibitem{BFK-ll} Barkai E, Fleurov V and Klafter J 2000 One dimensional stochastic L\'evy-Lorentz gas {\it Phys. Rev. {\rm E}} {\bf 61} 1164
\bibitem{BCLL-ll} Bianchi A, Cristadoro G, Lenci M and Ligab\`o M 2016 Random walks in a one-dimensional L\'evy random environment {\it J. Stat. Phys.} {\bf 163} 22
\bibitem{ST-ll} Sz\'asz D and T\'oth B 1984 Persistent random walks in a one-dimensional random environment {\it J. Stat. Phys.} {\bf 37} 27
\bibitem{BCV-ll} Burioni R, Caniparoli L and Vezzani A 2010 L\'evy walks and scaling in quenched disordered media {\it Phys. Rev. {\rm E}} {\bf 81}, 060101(R)
\bibitem{BBW-ll} Barthelemy P, Bertolotti J and Wiersma D S 2008 A L\'evy flight for light {\it Nature} {\bf 453} 495
\bibitem{BGA-ll} Beenakker C W J, Groth C W and Akhmerov A R 2009 Nonalgebraic length dependence of transmission through a chain of barriers with a L\'evy spacing distribution {\it Phys. Rev. {\rm B}} {\bf 79}, 024204
\bibitem{bigj} Vezzani A, Barkai E and Burioni R The single big jump principle in physical modelling arXiv:1804.02932v2
\bibitem{poll} Lewin L 1981 {\it Polylogarithms and associated functions} (New York: Elsevier)
\bibitem{Feller} Feller W 1966 {\it An Introduction to Probability Theory and Its Applications. Volume II} (New York: John Wiley)
\bibitem{Fla} Flajolet P and Sedgewick R 2008 {\it Analytic Combinatorics} (Cambridge: Cambridge University Press)
\bibitem{Ginn} Gillis J 1956 Centrally biased discrete random walk {\it Quart. J. Math.} {\bf 7} 144
\bibitem{IPZ} Ilyin V, Procaccia I and Zagorodny A 2013 Fokker-Planck equation with memory: the crossover
from ballistic to diffusive processes in many-particle
systems and incompressible media {\it Cond. Mat. Phys.} {\bf 16} 1
\bibitem{HPdiff} Hentschel H G E and Procaccia I 1984 Relative diffusion in turbulent media: the fractal dimension of clouds {\it Phys. Rev. {\rm A}} {\bf 29} 1984
\bibitem{OPdiff} O' Shaughnessy B and Procaccia I 1985 Analytical solutions for diffusion on fractal objects {\it Phys. Rev. Lett.} {\bf 54} 455
\bibitem{RGF} Regev F, Gr{\o}nbech-Jensen N and Farago O 2016 Isothermal Langevin dynamics  in systems with power-law spacially dependent friction {\it Phys. Rev. {\rm E}} {\bf 94} 012116
\bibitem{CC1} Cherstvy A G, Chechkin A V and Metzler R 2013 Anomalous diffusion and ergodicity breaking in heterogeneous diffusion processes {\it New J. Phys.} {\bf 15} 083039
\bibitem{CC2} Cherstvy A G and Metzler R 2015 Ergodicity breaking, ageing, and confinement in generalized diffusion processes with position and time dependent diffusivity {\it J. Stat. Mech.} P05010
\bibitem{SZS} Schmiedeberg M, Zaburdaeev V Y and Stark H 2009 On moments and scaling regimes in anomalous random walks {\it J. Stat. Mech.} P12020
\bibitem{CDP} Cipriani P, Denisov S and Politi A 2005 From anomalous energy diffusion to L\'evy walks and heat conductivity in one-dimensional systems {\it Phys. Rev. Lett.} {\bf 94} 244301
\bibitem{Vulp} Castiglione P, Mazzino A, Muratore-Ginanneschi P and Vulpiani A 1999 On strong anomalous diffusion {\it Physica {\rm D}} {\bf 134} 75
\end{thebibliography}
\end{document}